%% file: main.tex
\documentclass[conference]{IEEEtran}
\IEEEoverridecommandlockouts
\usepackage{cite}
\usepackage{amsmath,amssymb,amsfonts}
\usepackage{algorithmic}
\usepackage{graphicx}
\usepackage{textcomp}
\usepackage{xcolor}
\usepackage{caption}
\usepackage{url}
\usepackage{multirow}

\usepackage[caption=false,font=footnotesize,labelformat=simple]{subfig}

\begin{document}

\title{Touchdown on the Cloud:\\The impact of the Super Bowl on Cloud
}

\author{
\IEEEauthorblockN{Chen Wang}
\IEEEauthorblockA{IBM T. J. Watson Research Center\\
Yorktown Heights, NY， 10598 \\
Email: Chen.Wang1@ibm.com}
\and
\IEEEauthorblockN{Hyong Kim}
\IEEEauthorblockA{Carnegie Mellon University\\
Pittsburgh, PA 15213\\
Email: kim@ece.cmu.edu}}

\maketitle

\begin{abstract}
\input{abstract}
\end{abstract}


\input{introduction}
\input{relatedwork}
\input{datacollection}
\input{internet-impact}
\input{cloudvm-impact}
\input{sblivestreaming-impact}
\input{webservice-impact}
\input{conclusion}

\bibliographystyle{splncs04}
\bibliography{main}

\end{document}

%% file: abstract.tex
The Super Bowl is the world's biggest televised sporting event. We examine the impact of the increasing online activities during the Super Bowl of year 2015-2017 on various types of Cloud systems, including the Cloud infrastructure service, the Cloud content delivery networks for live streaming, and popular web services such as on-demand video streaming service and social network applications. We probe these systems from agents deployed around the world to compare their load variations and performance changes during and after the game. Through our studies of three consecutive years, we find that Super Bowl events have impacts on the  Cloud. However, the  current  Cloud  system  is still able to ensure capacity to cope with the challenges brought by traffic changes during such massive events.

%% file: introduction.tex
\section{Introduction}
The Super Bowl is the annual championship game of the National Football League (NFL), which is considered to be one of the largest national holidays, known as the ``Super Bowl Sunday'' in United States. It has been among the most-watched television broadcast every year. In particular, recent technological advances have significantly changed how people watch it. Nowadays, more and more people watch the game through live streaming services\cite{cha2008}. People may resort to the ``second screen'', such as tablets and smart phones, for an enhanced viewing experience. This is typically recognized as the ``social TV``, which allows people to take active roles in creating content by sharing their watching experience with others on social media such as Twitter and Facebook \cite{hanna2011, hill2012}. For example, while watching the Super Bowl, viewers may choose to tweet about the game \cite{han2017}, search the brands mentioned in ads \cite{lewis2013} and order the food, all conveniently through mobile apps. This creates strong incentives for companies to leverage online platforms and services to engage with them \cite{hanna2011}.

Understanding how increasing online activities during the Super Bowl may cause variations in the network traffic and the Cloud infrastructures remains under-explored. In general, Internet usage during the Super Bowl may get affected by how people watch the game. Traditionally, if the majority of viewers watch the game on cable TV, the Internet usage is expected to drop significantly during the game. To the contrary, if viewers increasingly watch the game via online streaming services, the transition would incur bursty traffic and significant amount of workload in Content Delivery Networks (CDNs) as well as the online streaming systems \footnote{One determinant is the availability of the live streaming for the Super Bowl, where the live content was legally streamed since 2012.}. Meanwhile, the Super Bowl may have divergent impact on different web services. For example, social media platforms where users tend to be more active during the game might experience much longer latencies than usual. However, viewers who were watching the game would not be able to consume video on demand services, such as Netflix and Hulu, which accounts for a big chunk of downstream traffic in usual days. Sandvine reported that during the 2012 Super Bowl game, overall Internet traffic was 20 percent lower compared to an average Sunday evening, while Netflix experienced a whopping 40 percent decrease in activity \cite{vb2012}.

In particular, as video streaming and web services continuously migrate to the Cloud for its reliability, supporting infrastructure and full suite of service, load in the Cloud infrastructure may also vary due to the changes in the demand. For example, Shazam, a leading media discovery app that helps people to identify media content around them, is deployed in Amazon Web Service (AWS). Throughout the 2012 Super Bowl event, Shazam had uplift in demand on its service for advertising campaign and chose the Cloud technology to ensure capacity for dramatic activity spikes \cite{shazam-aws}, thus it caused traffic spikes in AWS data centers at the same time. Therefore, it is imperative to understand how the Cloud infrastructure hosting different types of services may scale the resources to cope with challenges brought by traffic changes during such massive events.




In this paper, we aim at identifying the impact of the Super Bowl event on Internet, Cloud, CDN, as well as different types of web services, such as social networks and on-demand video streaming service. We do so by deploying hundreds of agents around the world to probe Cloud systems, including the virtual machines from three leading Cloud providers (AWS, Microsfot Azure and Google Cloud), the CDNs (from Akamai, Level3, Limelight) used for the live streaming, cache servers of popular social networks including Facebook and Twitter, and cache servers from Netflix. We obtain latency between agents and different systems in the Cloud infrastructure and compared these measurements among several periods in the same year: Super Bowl game time, halftime show time and the following Sunday nights, as well as across three consecutive years.

We have the following findings for the Super Bowl's impact on the Cloud networking, Cloud infrastructures and Cloud services. First, the Internet connection to Cloud at Super Bowl nights gets slower from 2015 to 2017 and gets slower than a normal Sunday night since 2016. Second, the Super Bowl games affect the internal networking of Cloud infrastructures. The average intra-data-center latency has increased more than 10 times in some Cloud data centers during the Super Bowl nights. However, most data centers still maintain the average intra-data-center latency within 10 ms. Moreover, all studied data-centers are able to maintain the variations of the inter-data-center latency within 5\%. Third, as of 2017, Super Bowl live streaming is provided by Fox, which is delivered through 3 CDNs, including Akamai, Level 3, and Limelight Networks. Our results indicate that performance of the Super Bowl live streaming varies among these CDN providers. However, within one provider, the difference between the Super Bowl live streaming and the Fox Sports Go live streaming for other sporting events is insignificant. Fourth, although Super Bowl game traffic may overload certain systems of web services, other systems become under-loaded. We show that it is actually faster to access Netflix and Facebook cache servers during the Super Bowl games,  whereas it is slower to access Twitter cache servers. Therefore, our findings suggest that although Super Bowl events have strong impacts on the Internet to Cloud, Cloud infrastructures and Cloud services, due to the increasing yet divergent online activities, the current Cloud system is able to rapidly ratchet up capacity to cope with the world’s biggest sporting event.

%% file: relatedwork.tex
\section{Related Work}
\subsection{Impact of Super Bowl on networking systems}
Overall, studies on the impact of the Super Bowl game on networking systems are very limited. Allen \textit{et al.} found unusual frequency disturbances in electricity network during the Super Bowl, due to the aggregated impact of synchronized television programming \cite{allen2016}. Jeffrey \textit{et al.} measured user behaviors and traffic demand of a large ISP's LTE cellular network at the Superdome stadium for Super Bowl in 2013. They found that streaming high-quality video at venues may represent a significant source of traffic. Nestor \textit{et al.} analyzed the BGP updates registered during Super Bowl 2016 to understand how it affected the Internet at BGP level \cite{morales2017did}. They found that an increase in the number of updates happened during key times when a large number of concurrent viewers connected to the game, thus the Internet was not ready to accommodate the potential streaming traffic generated by large events. In comparison, our work studies the impact of the Super Bowl game, yet on the Cloud infrastructures, the CDNs and web services in the Cloud.

\subsection{Impact of Super Bowl on social media}
By contrast, many studies have documented the impact of the Super Bowl on the social media. Han \textit{et al.} emphasized the viewer's social TV behaviors and engagement on social media during Super Bowl game for creating a pseudo-communal viewing experience \cite{han2017}. Hill \textit{et al.} performed the content analysis of tweet about products mentioned 2012 Super Bowl commercials and identified social media sentiment response   \cite{hill2012}. Lee \textit{et al} explored how Twitter users reproduced or contested the Super Bowl 2013 in reaction to a real-time televised broadcast \cite{lee2014sports}. Their macro-level analysis on twitter usage pattern gave insights about the twitter usage, the tweet instantaneity, the effects of the commercials in tweets, etc. Oh \textit{et al.} examined social media in relation to ad likeability ratings obtained from USA Today Ad Meter surrounding 2014 Super Bowl TV advertisements \cite{oh2015social}. They found that social media measures such as tweet volume and sentiment pertaining to Super Bowl were positively correlated with ad likeability ratings. Mukherjee and Jansen compared the correlations between the relative volume of 2015 Super Bowl commercials on Google search and that of posting on social media platforms, and found temporal trends of brand mentions and interplay between different channels \cite{mukherjee2015}. Instead, our work focuses on the impact of Super Bowl on the Cloud systems where social media platforms choose to host their services.

%% file: datacollection.tex
\section{Data Collection}
\subsection{Methodology}
We deployed hundreds of agents in PlanetLab cloud \cite{chun2003planetlab} around the world to probe various systems. Agents performed ICMP pings periodically to various types of systems and collect latency measurements. The endpoints being probed include the IP addresses of virtual machines (VMs) deployed in popular Clouds, the CDN host names used for Super Bowl live streaming, the host names of popular websites, and the host names of cache servers for web services. 

To study the variation of the network traffic and the load in the the cloud systems due to the Super Bowl game, we use the same set of agents to collect the latency measurements at the Super Bowl nights and at the following Sunday nights. We recorded the starting and ending time of all halftime shows and studied the impact of half time shows as they were special events during the Super Bowl. 

We collect measurements for Super Bowl games in the years of 2015-2017. As the available resources in PlanetLab and in the production Clouds changed over the years, we had slightly different settings over the years.

\subsection{Measurement of cloud VMs}
In 2015, we mainly studied the impact of the Super Bowl on the Internet traffic and the load on the cloud data-centers. We used 300 PlanetLab agents around the world to probe VMs every 5 minutes. Specifically, we started totally 32 VMs, 12 VMs in Google Cloud, 10 VMs in Microsoft Azure, and 8 VMs in AWS EC2. These VMs were deployed around the world in data-centers at different locations \footnote{In Google Cloud, there were 2 VMs in "us-central1-a" data-center in Iowa, 2 VMs in "us-central1-b" data center and 2 VMs in "us-central1-f" data center (both are in Oklahoma). In Europe, there were 2 VMs in "europe-west1-b" data center in Belgium and 2 VMs in "europe-west1-c" data center in Netherlands. In Asia, there were 6 VMs in 3 data centers ("asia-east1-a", "asia-east1-b" and "asia-east1-c") in Taiwan, 2 VMs per data center. In Microsoft Azure Cloud, we created 10 VMs in 10 data centers at 9 locations. These data centers are "East US" and "East US 2" in Virginia, "Central US" in Iowa, "South Central US" in Texas, "West US" in California, "North Europe" in Ireland, "West Europe" in Netherlands, "East Asia" in Hong Kong, "Southeast Asia" in Singapore, and "Japan West" in Osaka Prefecture. In AWS, we deployed 8 VMs in 8 data centers. These data centers are at North Virginia, Oregon, California, Ireland, Singapore, Tokyo, Sydney, and Sao Paulo.}.

In 2016, in addition to the measurement of network latencies between Cloud VMs and external PlanetLab agents, we also measured inter-data center and intra-data center network latencies by agents deployed in VMs in the Clouds. We probed VMs only in US as we noticed from 2015 that the Super Bowl game had few impact on the systems out of US. In AWS and Azure Cloud, we started 2 VMs per each data center in the east coast. In Google Cloud, we noticed there were 3 new data centers available in US in 2016\footnote{They were "us-east1-a", "us-east1-b", "us-east1-f" and "us-central1-c".}, so we deployed 14 VMs in 7 data centers, with 2 VMs per data center.

In 2017, we found that the Google Cloud had 3 more data centers provided in US west coast. As we had gained insights about the inter/intra-data center latencies from 2016, we only chose 3 data centers in 3 regions and deployed 2 VMs per data center for comparison of Cloud infrastructure service over years \footnote{In AWS, we deployed 1 VM in US east and 1 VM in US west. In Microsoft Azure, we deployed 3 VMs in 3 data centers in the east US region, the central US region and the west US region respectively.}. 

\subsection{Measurement of CDNs for Super Bowl live streaming}
Since 2016, we collected network measurements to servers hosting the website of live streaming. In 2016, CBS Sports offered live streaming for Super Bowl 50. We probed the host name of \url{www.cbssports.com} for CBS Sports website every 1 minute.

In 2017, Fox Sports GO offered the live streaming for Super Bowl LI. However, the video content were actually cached and distributed via multiple CDNs. We found the following host names of these CDNs that served as cache servers for Super Bowl live streaming service. We probed these cache servers every 1 minute from 100 PlanetLab agents. 
\begin{table*}[!h]
\centering
\begin{tabular}{c|c}
\hline
\textbf{CDN Provider} & \textbf{Host names of CDN cache servers} \\
& \textbf{for Super Bowl live streaming service} \\
\hline
\multirow{2}{*}{\textbf{Level 3}} & \url{hlsevent-13c.med1.foxsportsgo.com} \\ 
& \url{hlsevent-13c.med2.foxsportsgo.com} \\
\hline
\multirow{2}{*}{\textbf{Limelight Networks}} & \url{hlsevent-llc.med1.foxsportsgo.com} \\ 
& \url{hlsevent-llc.med2.foxsportsgo.com} \\
\hline
\multirow{2}{*}{\textbf{Akamai}} & \url{hlsevent-akc.med1.foxsportsgo.com} \\ 
& \url{hlsevent-akc.med2.foxsportsgo.com} \\
\hline
\end{tabular}
\caption{Host names of CDN cache servers for Super Bowl LI live streaming services}
\label{tbl:sb_2017_cdn_host}
\end{table*}

\subsection{Measurement of web applications}
\subsubsection{Social networks}
There are two possible cases that the Super Bowl could impact the social network usage.  One case is that there would be boosting usage in social networks as people have hot discussions about NFL during Super Bowls. The other case is that people would be so focused on watching the game and would not pull the updates from social networks so often as they usually do.  In order to understand which case really applies, we probed Twitter and Facebook cache servers. We measured the latencies from PlanetLab agents to their cache servers at Super Bowl night and the following Sunday night. Facebook and Twitter cache their images in their own CDNs as images incur the most of the traffic in social network applications. We also wonder if the Super Bowl would affect the performance of their CDN cache servers. Therefore, we probed the host names of their CDN servers for Twitter (\url{pbs.twimg.com}) and Facebook (\url{fbcdn.net}). 

\subsubsection{On demand video streaming service}
Not all applications become popular during Super Bowl. Some applications may have fewer workload at Super Bowl nights. As more than 70\% of Internet traffic are Internet video traffic, it is reasonable to infer that most users watch movies or TV shows online for their entertainments at weekend nights \cite{cisco}. 
Popular on-demand video streaming services include Netflix, Hulu, YouTube, and Amazon prime video. However, at Super Bowl nights, they may watch the Super Bowl instead. We probed hosts of Netflix video streaming in 2016 and 2017 to study the impact of Super Bowl games on Netflix services. We found an example host name for Netflix video streaming service from the chunk requests we captured when playing Netflix videos. The host names we probed at was \url{netflix753.as.nflximg.com.edgesuite.net} in 2016 \footnote{We found that the host name belonged to Akamai.}.

%% file: internet-impact.tex
\section{Impact of Super Bowl on the Internet to Cloud}
\label{sec:sb_on_internet}
\begin{figure*}[!htb]
\centering
   \includegraphics[width=0.9\textwidth]{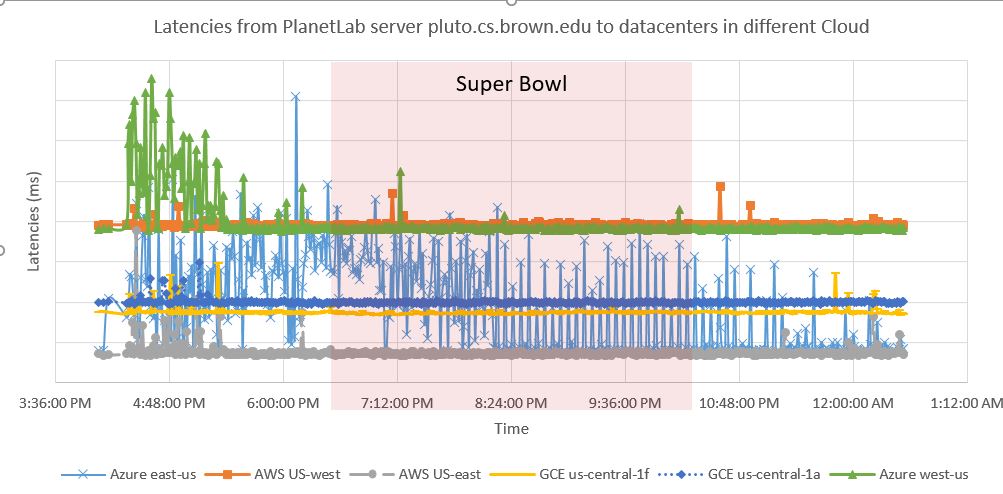}
   \caption{Latencies probed from PlanetLab agent "pluto.cs.brown.edu" to all Cloud VMs in US during Super Bowl XLIX}
   \label{fig:sb_lat_2015}
\end{figure*}

We first give a brief view on how the latency changes due to the Super Bowl game. We draw the latencies from one PlanetLab agent to VMs in data centers from different Cloud providers in Figure \ref{fig:sb_lat_2015}. We extend the time range to include 4 hours before Super Bowl XLIX and 4 hours after the game. Thus, we show how the probed latencies varied before, during and after the game. Overall, we observe more variations before the game.

\subsection{Time periods for comparison}
As the game was in the prime time on Sunday, it would be not be fair to compare latencies in the afternoon or in the midnight with the latencies in the prime time. Therefore, we compare the latencies at Super Bowl game night and the following Sunday night to control the impact of other variables, such as the day in a week or the time in a day, etc. As we used agents in PlanetLab to probe the VMs in various Cloud providers, the Internet traffic variations due to the Super Bowl game can be reflected in the latencies. As the half time show is also an eye catching event, we also study it as a separate time period. In summary, we compare the latency measurement in 3 time periods, \textit{superbowl} denoting the game period, \textit{halftime} denoting the half time show period, and \textit{postsuperbowl} denoting the following Sunday night after the super Bowl week.

\subsection{Latency from PlanetLab agents to Cloud virtual machines}
\begin{figure*}[htp]
\centering
    \begin{minipage}[b]{\textwidth}
        \subfloat[Super Bowl XLIX in 2015]{
            \includegraphics[width=0.32\textwidth]{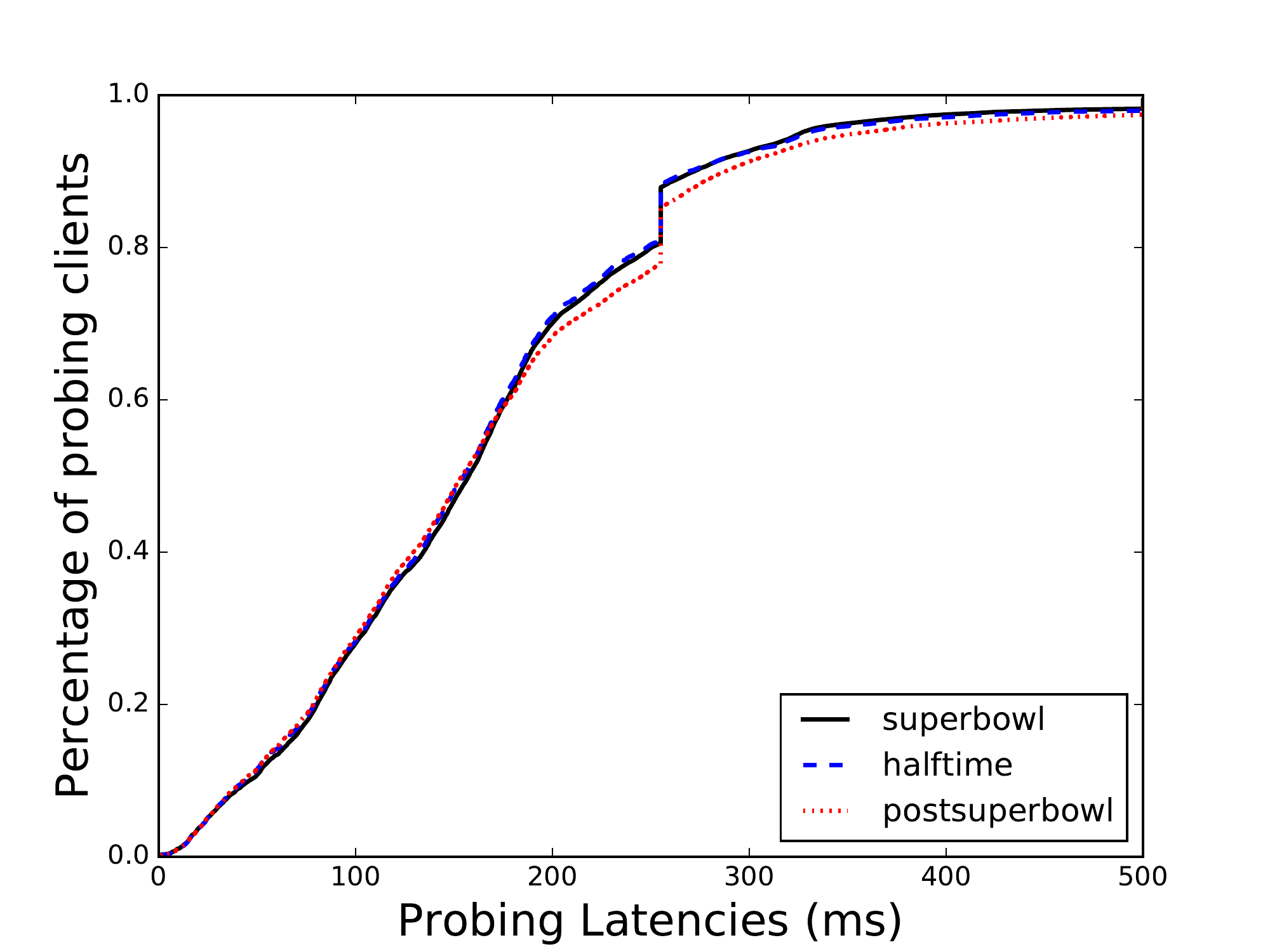}
        }
        \subfloat[Super Bowl 50 in 2016]{
            \includegraphics[width=0.32\textwidth]{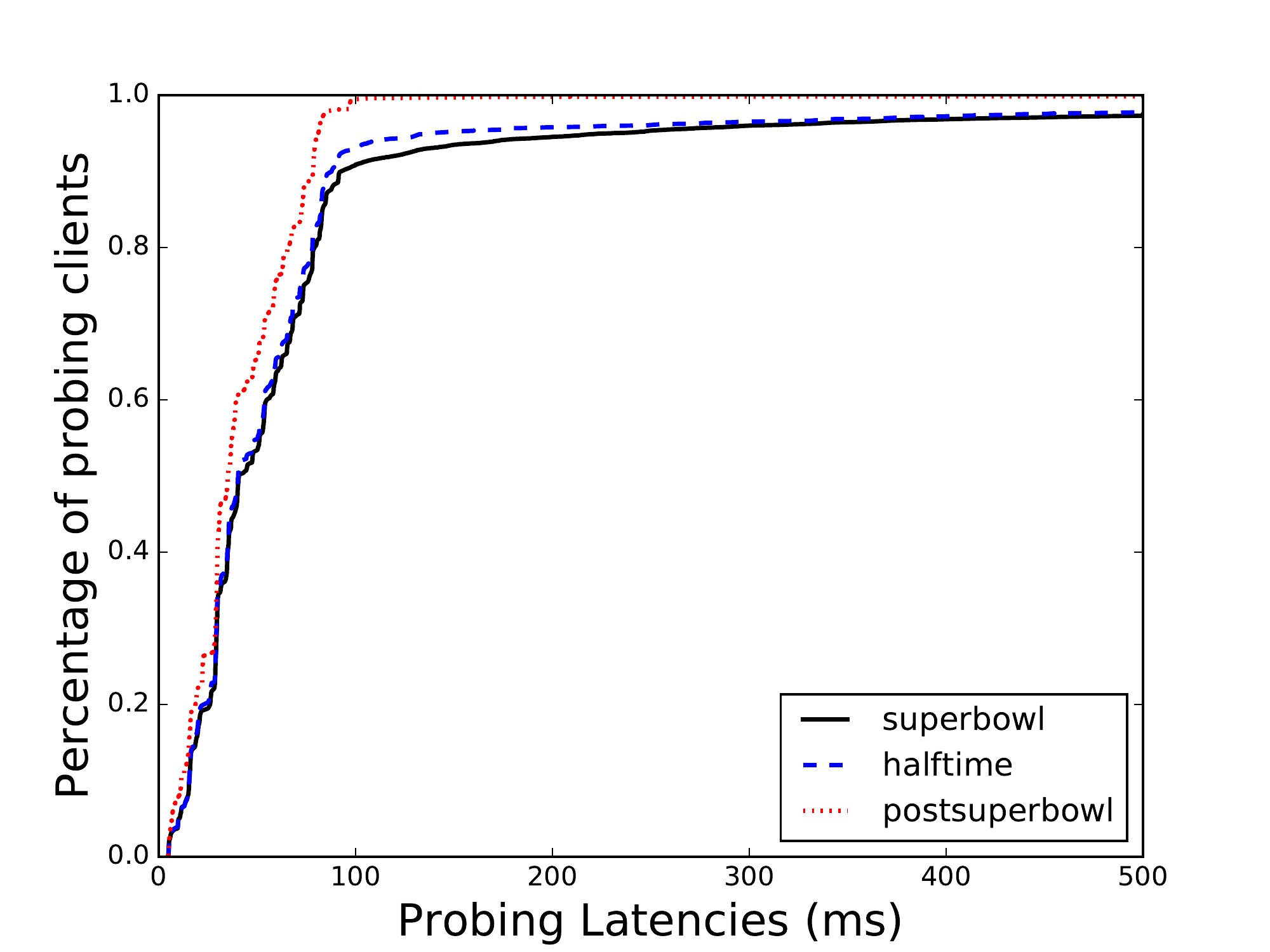}
        }
        \subfloat[Super Bowl LI in 2017]{
            \includegraphics[width=0.32\textwidth]{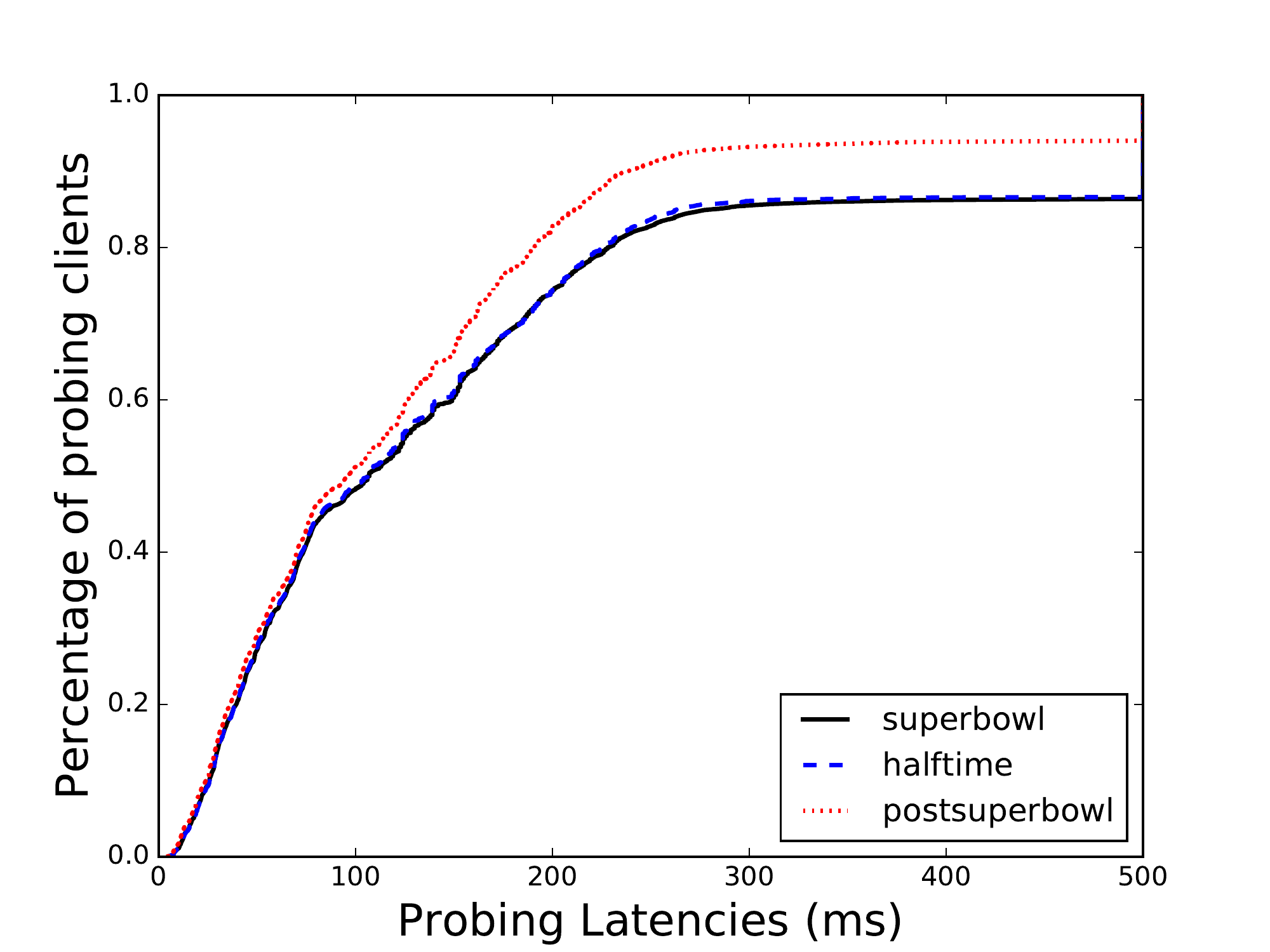}
        }
    \end{minipage}
    \caption{The CDF of latencies from PlanetLab agents to all Cloud VMs in US from 2015 to 2017}
    \label{fig:sb_period_cmp_3yrs}
\end{figure*}

We first compare the cumulative distribution of latencies from all PlanetLab agents to all Cloud VMs in US among above 3 time periods. Figure \ref{fig:sb_period_cmp_3yrs} (a), (b), and (c) show the data in 2015, 2016 and 2017 respectively. In Figure \ref{fig:sb_period_cmp_3yrs} (a), we show that the latency distribution in 3 time periods were similar, yet the latencies in \textit{superbowl} period were equal or shorter than the latencies during the \textit{postsuperbowl} period. Thus we infer that the Internet connection to the Cloud during the game period were faster than the following Sunday night in 2015.
However, we got completely opposite insights from the latency distribution comparison in 2016 and 2017. Figure \ref{fig:sb_period_cmp_3yrs} (b) shows the latencies from 10 closest PlanetLab agents in US to each Cloud VM in US in 2016. To reduce the volume of ping traffic during the Super Bowl, we only kept latency measurements from 10 closest PlanetLab agents of each VM in 2016. The overall latencies were much shorter with 90th percentile latencies less than 100 ms. We also observe a small percentage of latencies were at 500 ms. These latencies actually got timeout at the time of probing \footnote{The 500 ms is the threshold we set to prevent the large number of pending sessions.}. We see clearly that in 2016, the latencies during \textit{postsuperbowl} period were smaller than the \textit{superbowl} at almost all percentiles. It indicated that the Internet connections to cloud VMs in US were slower during Super Bowl 50 than the following Sunday nights.
\begin{table*}[!htb]
\centering
\begin{tabular}{c|c|c|c|c|c|c}
\hline
\multirow{2}{*}{\textbf{Year}} & \multicolumn{2}{c|}{\textbf{\textit{superbowl} (ms)}} & \multicolumn{2}{c|}{\textbf{\textit{halftime} (ms)}} & \multicolumn{2}{c}{\textbf{\textit{postsuperbowl} (ms)}} \\ \cline{2-7}
& \textbf{Mean} & \textbf{STD} & \textbf{Mean} & \textbf{STD} & \textbf{Mean} & \textbf{STD} \\ \hline
\textbf{2015} & 157.86 & 86.41 & 155.88 & 86.54 & 158.64 & 91.06 \\ 
\textbf{2016} & 55.96 & 54.81 & 52.84 & 50.52 & 40.99 & 24.77 \\
\textbf{2017} & 105.26 & 76.01 & 104.12 & 74.31 & 104.52 & 73.57 \\ \hline
\end{tabular}
\caption{Mean and STD of latencies in 3 periods in 2015, 2016 and 2017.}
\label{tbl:lat_stats_internet_3yrs}
\end{table*}
We observe similar latency distributions in 2017. Compared to 2016, the gap between 90th percentile latency in the \textit{superbowl} and the \textit{postsuperbowl} periods was larger in 2017. We observe that there were more than 10\% of latencies greater than 500 ms in 2017 during the \textit{superbowl} period. These were actually timeout requests. From 2015 to 2017, we observe an increasing gap between the latency distribution in \textit{superbowl} and the latency distribution in the \textit{postsuperbowl}. It indicate that due to the Super Bowl the Internet connections to cloud VMs were slowing down and such an impact was increasing over 3 years. As the number of agents we used for experiments were not the same over the years, the mean of latencies we observed in Table \ref{tbl:lat_stats_internet_3yrs} varied\footnote{Timeout probes were considered as having 500 ms latencies.}.  However, the increasing gaps due to Super Bowl games over the years indicate an increasing Internet traffic during the game, which might be incurred by the live streaming traffic, the social network activities, or various web application related to Super Bowl.

%% file: cloudvm-impact.tex
\section{Impact of Super Bowl on Cloud infrastructure service}
\label{sec:sb_impact_cloud}
To study the impact of Super Bowl on the Cloud infrastructure as a service (IaaS), we compare the intra and inter data center network latencies among three periods in 2016. The study is based on the assumption that the latencies observed by VMs in the Cloud can be affected by the load variations in Cloud infrastructures due to Super Bowl games.

\subsection{Intra-data center latencies}
Similar to Section \ref{sec:sb_on_internet}, we compare the network latencies between VMs in the Cloud in three periods: the \textit{superbowl}, the \textit{halftime} and the \textit{postsuperbowl}. We assume that the intra-data center observed by idling VMs in the Cloud would be higher if the workload in the data center is higher.

\begin{table*}[!htb]
\centering
\begin{tabular}{c|c|c|c|c|c}
\hline
\textbf{Data Center}  & \textbf{Period (ms)} & \textbf{Mean (ms)} & \textbf{Min (ms)} & \textbf{Max (ms)} & \textbf{STD (ms)} \\ \hline
\multirow{3}{*}{\textbf{Virginia}} & \textbf{\textit{superbowl}} & 2.1789 & 1.0500 & 20.00 & 2.5533 \\ 
& \textbf{\textit{halftime}} & 1.7113 & 1.1100 & 4.13 & 0.7104 \\ 
& \textbf{\textit{postsuperbowl}} & 1.6060 & 1.1600 & 10.40 & 1.1059 \\ \hline
\multirow{3}{*}{\textbf{California}} & \textbf{\textit{superbowl}} & 0.9606 & 0.4890 & 8.9700 & 1.1540 \\ 
& \textbf{\textit{halftime}} & 1.3134 & 0.4890 & 7.2900 & 1.7733 \\ 
& \textbf{\textit{postsuperbowl}} & 0.8673 & 0.6180 & 6.5900 & 0.5731 \\ \hline
\end{tabular}
\caption{Statistics of intra-data center latencies (ms) for data centers in AWS during 3 periods.}
\label{tbl:aws_intra_dc_lats_3yrs}
\end{table*}

Table \ref{tbl:aws_intra_dc_lats_3yrs} shows the statistics data of intra-data center latencies in AWS, collected during the \textit{superbowl}, the \textit{halftime} and the \textit{postsuperbowl} periods in 2016. Each row showed the statistics of all latencies obtained between two VMs deployed in the same data center in the specified period. We studied two data centers, one in Virginia and the other in California. 

First, we observe the latencies in different periods in both AWS data centers. Comparing the data in the \textit{superbowl} and the \textit{postsuperbowl} periods, we observe an increase in both the average and the standard deviation of intra-data center latencies due to the game. As the game was the only factor we controlled and the increase in the latencies occured in both data centers, we inferred that such latency variations were due to the Super Bowl game. Possible reasons can be high load in the data centers or high interference in the VMs being probed during the game.

Second, we compared the average and the standard deviation of intra-data center latencies between the \textit{superbowl} and the \textit{halftime} periods. As both were at the Super Bowl game night, the factor we controlled was the half time show event. We find that the location of the data center also affect the intra-data center latencies. In the data center in California, where the Super Bowl 50 was held, the latencies during the half time show had higher average and deviations than the game period. In the data center in Virginia, which is far from the game location, the latencies in the game period were higher and more varied than the half time show. One possible guess can be that there were more Super Bowl related traffic received in the data center closer to the game location during the Halftime show. Comparing the period with the minimum mean latency (\textit{postsuperbowl}) with the period with the maximum mean latency (the \textit{halftime} period in California and the \textit{superbowl} period in Virginia), we find that the impact of Super Bowl on the intra data center latencies varied across locations. In in Virginia, it was on average 35.67\%. In California, it was on average 51.43\%. Though the absolute increase was small, we can see that the Super Bowl had a higher impact on the system closer to the game location.
\begin{table*}[htb]
\centering
\begin{tabular}{c|c|c|c|c|c}
\hline
\textbf{Data Center}  & \textbf{Period (ms)} & \textbf{Mean (ms)} & \textbf{Min (ms)} & \textbf{Max (ms)} & \textbf{STD (ms)} \\ \hline
\multirow{3}{*}{\textbf{Virginia}} & \textbf{\textit{superbowl}} & 31.2613 & 1.2500 & 86.5000 & 25.9114 \\ 
& \textbf{\textit{halftime}} & 30.4177 & 1.2800 & 86.5000 & 27.5301 \\ 
& \textbf{\textit{postsuperbowl}} & 1.9520 & 0.3660 & 25.8000 & 2.8201 \\ \hline
\multirow{3}{*}{\textbf{California}} & \textbf{\textit{superbowl}} & 5.5338 & 0.5680 & 49.600 & 8.2015 \\ 
& \textbf{\textit{halftime}} & 2.9036 & 0.6800 & 20.700 & 4.2538 \\ 
& \textbf{\textit{postsuperbowl}} & 1.6285 & 0.6210 & 16.800 & 2.0281 \\ \hline
\end{tabular}
\caption{Statistics of intra-data center latencies (ms) for data centers in Azure during 3 periods.}
\label{tbl:azure_intra_dc_lats_3yrs}
\end{table*}

Table \ref{tbl:azure_intra_dc_lats_3yrs} shows the statistics of intra-data latencies for data centers in Azure. Similarly, we find that the Super Bowl game had increased the average and variation of intra-data center latencies in both Virginia and California in Azure. However, the impact of the half time show over the game was different. In both data centers, we observe that the average and variation of intra-data center latencies were higher in the \textit{superbowl} period than in the \textit{halftime} period. In Azure Cloud, we observed a much stronger impact of the Super Bowl as the mean intra-data center latencies increased up to more than 16x in the Virginia data center during the game. 
\begin{table*}[htb]
\centering
\begin{tabular}{c|c|c|c|c|c}
\hline
\textbf{Data Center} & \textbf{Period (ms)} & \textbf{Mean (ms)} & \textbf{Min (ms)} & \textbf{Max (ms)} & \textbf{STD (ms)} \\ \hline
\multirow{3}{*}{\textbf{South Carolina}} & \textbf{\textit{superbowl}} & 1.3829 & 0.8440 & 16.800 & 1.636 \\ 
& \textbf{\textit{halftime}} & 1.1099 & 0.9280 & 1.300 & 0.0866 \\ 
& \textbf{\textit{postsbowl}} & 1.1185 & 0.4850 & 8.80 & 0.5897 \\ \hline
\multirow{3}{*}{\textbf{Iowa}} & \textbf{\textit{superbowl}} & 1.3183 & 0.6210 & 14.7000 & 1.2636 \\ 
& \textbf{\textit{halftime}} & 1.1256 & 0.9660 & 1.3300 & 0.0948 \\ 
& \textbf{\textit{postsbowl}} & 2.0054 & 0.6490 & 92.3000 & 8.4779 \\ \hline
\end{tabular}
\caption{Statistics of intra-data center latencies (ms) for data centers in Google Cloud during 3 periods.}
\label{tbl:google_intra_dc_lats_3yrs}
\end{table*}
In Table \ref{tbl:google_intra_dc_lats_3yrs}, we compared the intra-data center latencies for two Google Cloud data centers. They were at South Carolina and Iowa as there were only data centers in the central US available when we did our experiments. By comparing the \textit{superbowl} period with the \textit{postsuperbowl} period, we find the intra-data center latencies increased in the South Carolina data center but decreased in the data center in Iowa due to the superbowl. It is very different from all other data center we observed. Possible reason could be: 1) there might be fewer Super Bowl related applications in Google Iowa data center at that time; 2) users in Iowa may have less Internet activities during the Super Bowl game. We also observe that the increase of latencies due to varied traffic on average was 23.64\% in South Carolina data center.

Overall, the average intra-data center latencies studied in all above data centers were below 10 ms and it showed that though the impact of Super Bowl game existed, it did not significantly degrade the network performance of VMs. 
\subsection{Inter-data center latencies}
We then compared the inter-data center network latencies among the three periods. Similarly, we controlled the impact of the Super Bowl game and the half time show. Cloud providers might have their own network infrastructures for their inter-data center networking. They might also peer with other Internet Service Providers to transfer data among data centers. If they peer with others for the inter-data center networking, their latencies can be affected by the Internet traffic. Otherwise, their latencies are only affected by their internal load.

We first studied the AWS Cloud. As AWS deployed all data centers in the east and west coasts in US, we compared the inter-data center latencies between VMs in two data centers, "us-east1" and "us-west1", as shown in Table \ref{tbl:aws_inter_dc_lats}. In AWS, the impact of the Super Bowl game on the inter-data center networking was similar as what we observed in the intra-data center latencies. The AWS inter-data center latencies were on average higher at the Super Bowl night than the following Sunday night. Information about how AWS connected their data centers was not available to us, however, Super Bowl did change the inter-data center latencies. 
\begin{table}[htb]
\centering
\begin{tabular}{c|c|c|c|c}
\hline
\textbf{Period} & \textbf{Mean} & \textbf{Min} & \textbf{Max} & \textbf{STD}\\ \hline
\textbf{\textit{superbowl}} & 76.5517 & 71.7000 & 88.8000 & 2.0462 \\
\textbf{\textit{halftime}} & 76.6033 & 71.7000 & 83.8000 & 2.0377 \\
\textbf{\textit{postsuperbowl}} & 74.2395 & 71.6000 & 99.8000 & 3.1515 \\ \hline
\end{tabular}
\caption{Statistics of inter-data center latencies (ms) for data centers in AWS during 3 periods.}
\label{tbl:aws_inter_dc_lats}
\end{table}

As shown in Table \ref{tbl:azure_inter_dc_lats}, Azure Cloud also had higher inter-data center latencies during the Super Bowl. We observed that the increase of mean latencies due to the Super Bowl was less than 5\%.  The maximum inter-data center latency was around 101 ms. We can see that though there was small impact of Super Bowl on the inter-data center networking but it was hardly noticeable. 
\begin{table}[htb]
\centering
\begin{tabular}{c|c|c|c|c}
\hline
\textbf{Period} & \textbf{Mean} & \textbf{Min} & \textbf{Max} & \textbf{STD} \\ \hline
\textbf{\textit{superbowl}} & 69.8033 & 65.6000 & 101.0000 & 7.1997 \\ 
\textbf{\textit{halftime}} & 67.8133 & 65.7000 & 96.4000 & 6.1210 \\ 
\textbf{\textit{postsuperbowl}} & 66.8713 & 65.3000 & 88.7000 & 2.2697 \\ \hline
\end{tabular}
\caption{Statistics of the inter-data center latencies  (ms) between Azure "us-east1" to "us-west1" during 3 periods.}
\label{tbl:azure_inter_dc_lats}
\end{table}

Table \ref{tbl:google_inter_dc_lats} compared the inter-data center latencies in Google Cloud. The statistics showed similar gap in latencies between the \textit{superbowl} and the \textit{postsuperbowl} periods in two data centers. The latencies in the \textit{superbowl} period were also higher than the \textit{postsuperbowl} period. Similarly, the increase was less than 1\%.

\begin{table}[htb]
\centering
\begin{tabular}{c|c|c|c|c}
\hline
\textbf{Period} & \textbf{Mean} & \textbf{Min} & \textbf{Max} & \textbf{STD} \\ \hline
\textbf{\textit{superbowl}} & 37.3467 & 36.6000 & 52.2000 & 1.6775 \\ 
\textbf{\textit{halftime}} & 37.0833 & 36.8000 & 37.3000 & 0.1234 \\ 
\textbf{\textit{postsuperbowl}} & 37.0749 & 65.3000 & 88.7000 & 2.2697 \\ \hline
\end{tabular}
\caption{Statistics of the inter-data center latencies  (ms) between Google "us-central-1" to "us-central-4" data centers during 3 Super Bowl periods.}
\label{tbl:google_inter_dc_lats}
\end{table}

Overall, we observed that the Super Bowl indeed had an influence on the inter-data center networks in all Cloud providers.  However, the impact on the inter-data center latencies on average were little and can be negligible.

%% file: sblivestreaming-impact.tex
\section{Super Bowl live streaming}
\subsection{Super Bowl live streaming website}
We study the latencies from agents to Super Bowl live streaming websites in 2016. We deployed 10 PlanetLab agents probing the CBS Sports website host \url{www.cbssports.com}.

\begin{figure}[!htb]
   \includegraphics[width=0.5\textwidth]{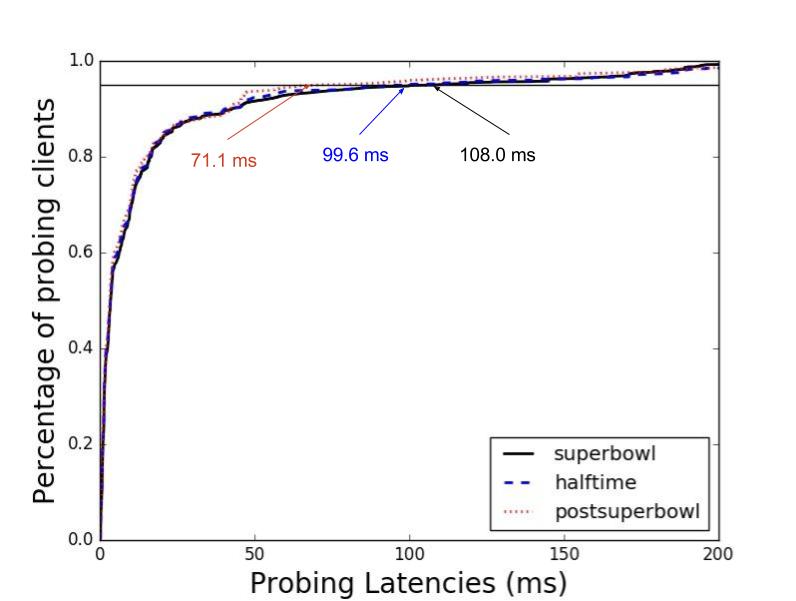}
   \caption{The CDF of probing latencies from PlanetLab agents to Super Bowl live streaming web site in 2016}
   \label{fig:livestreaming_website_2016_lat_cdf}
\end{figure}

We collected latency measurement during the \textit{superbowl}, the \textit{halftime} and the \textit{postsuperbowl} periods. We draw the cumulative distribution of all latency measurements in Figure \ref{fig:livestreaming_website_2016_lat_cdf}. The CDF curves show that the latencies among three different periods were similar. The top 95th percentile latency during the \textit{postsuperbowl} period is 71.1 ms, which was lower than the 95th percentile latency in the \textit{superbowl} period (108.0 ms) and the \textit{halftime} period (99.6 ms). We observe that the latency increase due to the Super Bowl and the halftime show only affect less than 10\% of latencies. The increase in the 90th percentile latency due to Super Bowl was around 30 ms. Compared to the intra and inter data center latencies we studied in Section \ref{sec:sb_impact_cloud}, 30ms increase was big. However, considering how bursty the load would be on the live streaming website during the Super Bowl game, we believed the system had scaled well. 
\vspace{2em}
\subsection{Content Delivery Networks for Super Bowl live streaming service}
In 2017, Fox Sports Go website offered the Super Bowl live streaming. By capturing and parsing the packets of the live streaming videos, we discovered three content delivery networks used for caching the Super Bowl live streaming videos. They were Akamai, Level 3, and LimeLight networks. We used 100 PlanetLab agents in US to probe the host names of cache services that delivered the Super Bowl videos, as shown in Table \ref{tbl:sb_2017_cdn_host}.

\begin{figure}[!htb]
   \includegraphics[width=0.5\textwidth]{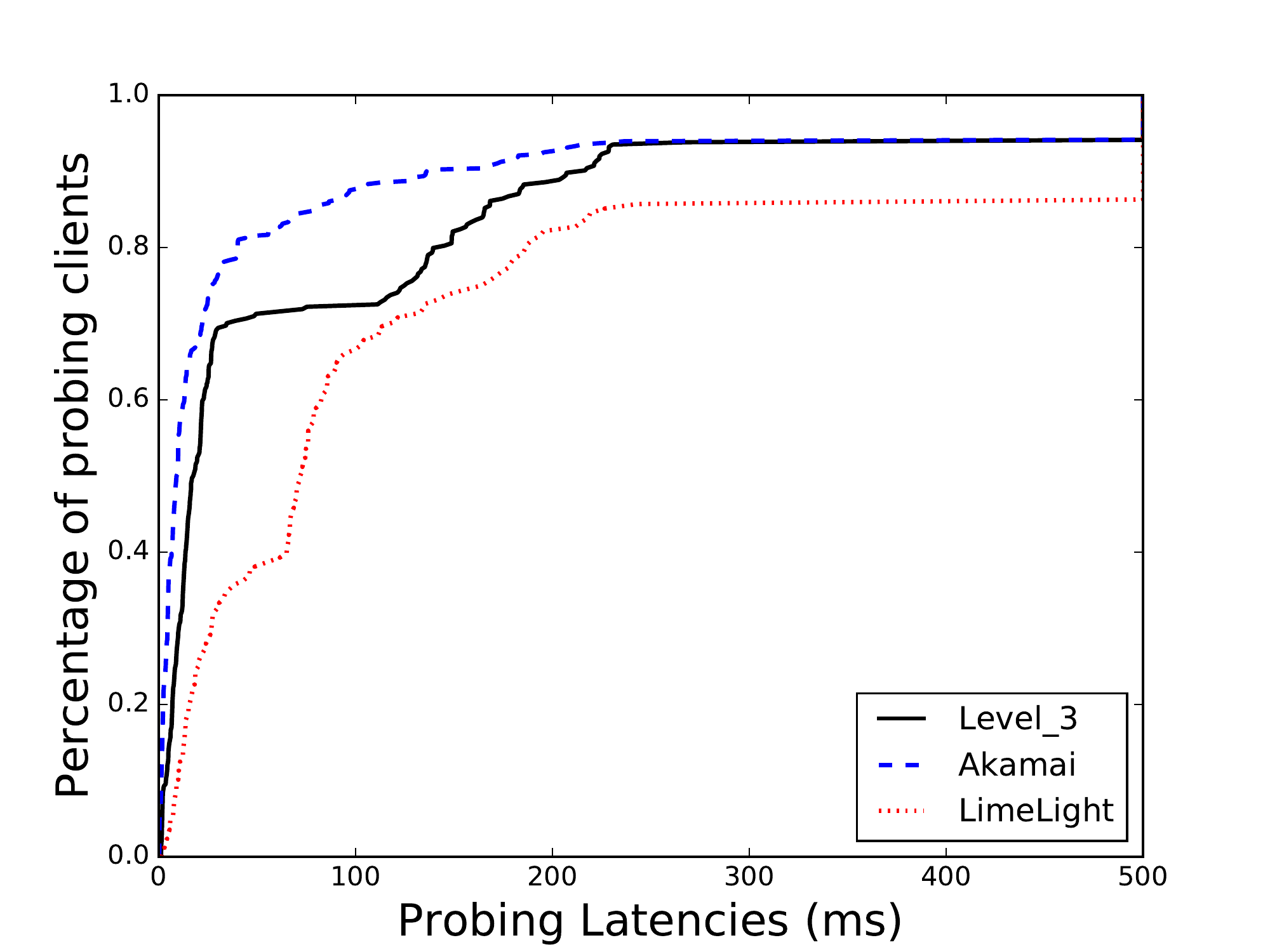}
   \caption{The CDF of probing latencies from PlanetLab agents to multiple CDNs that cached Super Bowl live streaming videos in 2017}
   \label{fig:livestreaming_sb_CDN_lat_cdf_2017}
\end{figure}

In Figure \ref{fig:livestreaming_sb_CDN_lat_cdf_2017}, we first compare the overall latencies to three CDN systems at the Super Bowl night. We observe that during the Super Bowl, PlanetLab agents had lower latencies to Akamai than the Level 3 CDN system. These agents also had lower latencies to Level 3 CDN than the LimeLight CDN.

In order to see if CDN systems had special resource provisioning set up for the bursty user demand of the Super Bowl live streaming, we also captured and parsed the packets from Fox Sports Go Live streaming and discovered the CDN hosts for it, as shown in Table \ref{tbl:fox_go_live_streaming_CDN_hosts}. We only discovered host names of cache servers in Level 3 and Akamai. We can see that Fox Sports Go indeed set up additional CDN resources to deliver Super Bowl live videos. 

\begin{table*}[!htb]
\centering
\begin{tabular}{c|c}
\hline
\textbf{CDN Provider} & \textbf{Host names of CDN cache servers for Fox Sports Go Live Streaming service} \\
\hline
\multirow{2}{*}{\textbf{Level 3}} & \url{hlslinear-l3c.med1.foxsportsgo.com} \\ 
& \url{hlslinear-l3c.med2.foxsportsgo.com} \\
\hline
\multirow{4}{*}{\textbf{Akamai}} & \url{hlslinear-akc.med1.foxsportsgo.com} \\ 
& \url{hlslinear-akc.med2.foxsportsgo.com} \\ 
& \url{hlsremote-akc1.us1.foxsportsgo.com} \\ 
& \url{hlsremote-akc1.us2.foxsportsgo.com} \\
\hline
\end{tabular}
\caption{Host names of CDN cache servers for Fox Sports Go Live Streaming service}
\label{tbl:fox_go_live_streaming_CDN_hosts}
\end{table*}

\begin{figure*}[!ht]
\centering
    \begin{minipage}[b]{\textwidth}
        \subfloat[Level 3]{
            \includegraphics[width=0.48\textwidth]{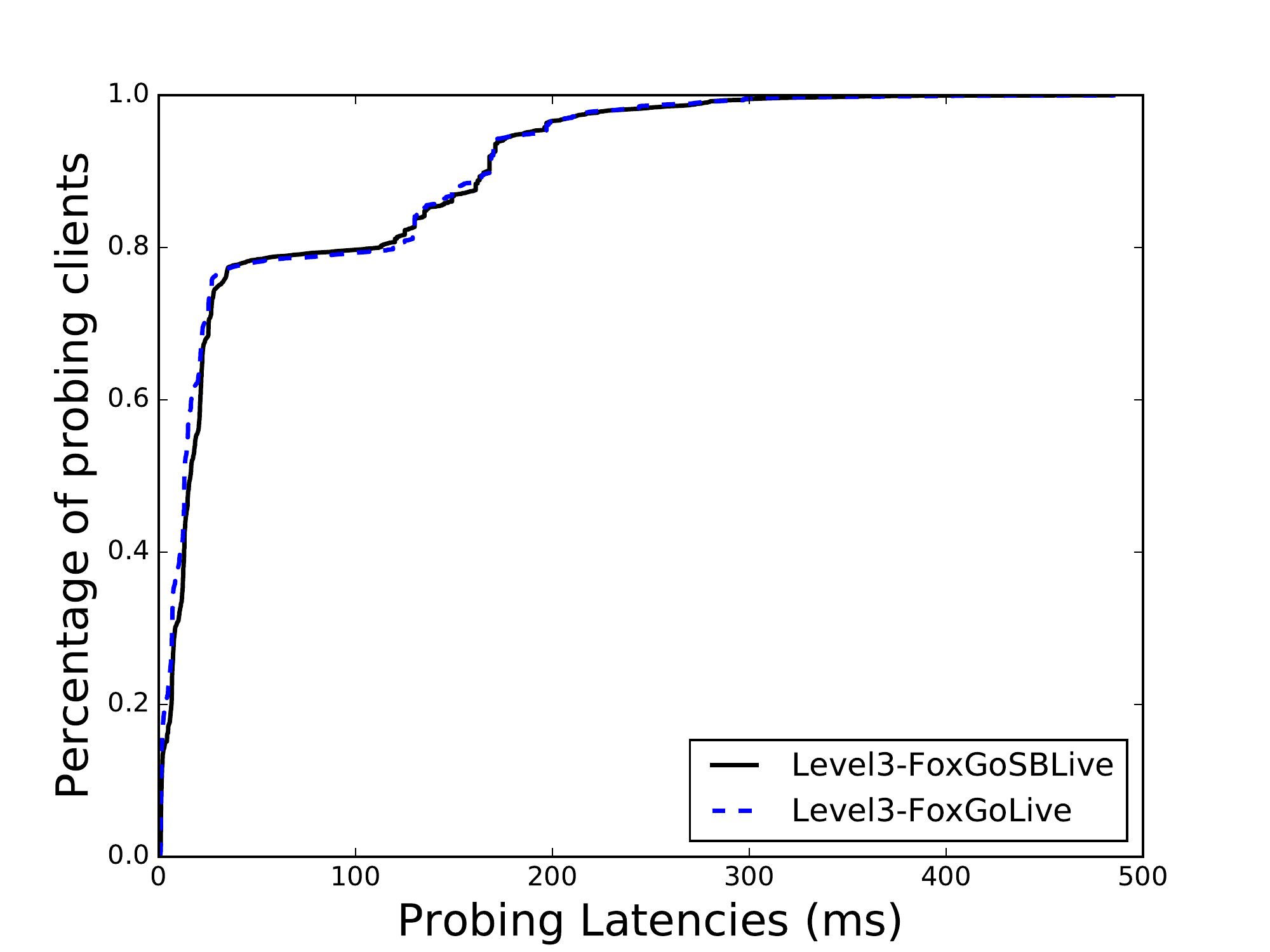}
        }
        \subfloat[Akamai]{
            \includegraphics[width=0.48\textwidth]{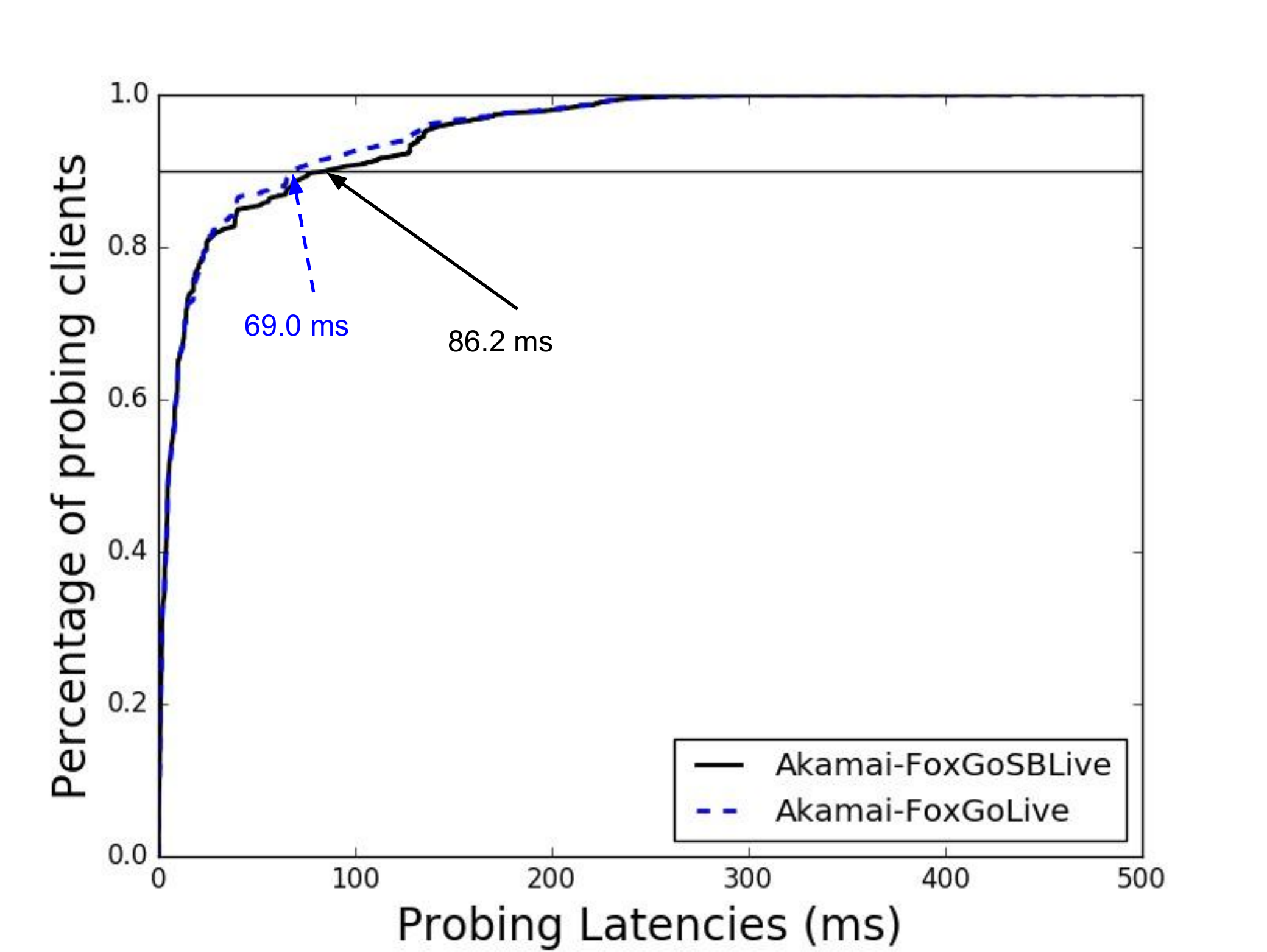}
        }
    \end{minipage}
    \caption{The CDF of latencies from PlanetLab agents to CDN hosts of Super Bowl live streaming and to CDN hosts of Fox Sports Go Live streaming during Super Bowl LI}
    \label{fig:foxgo_vs_sb_2017}
\end{figure*}
We then compared the latencies to Super Bowl live streaming CDN hosts with the latencies to Fox Sports Go live streaming CDN hosts in Akamai and level 3 in Figure \ref{fig:foxgo_vs_sb_2017}. Figure \ref{fig:foxgo_vs_sb_2017} (a) shows that there were no significant difference between latencies to the cache servers of Super Bowl live streaming and the cache servers of Fox Sports Go live streaming hosts. It indicated that during the Super Bowl game, the Level 3 CDN system had scaled resources provisioning well and appropriately handled the bursty demand of the Super Bowl live streaming. We also compared the latencies to cache servers in Akamai in Figure \ref{fig:foxgo_vs_sb_2017} (b). Similarly, PlanetLab agents had similar latencies to different Akamai cache servers. If we observe the latencies at the 90th percentile, there was slightly higher latency to the cache servers of the Super Bowl live streaming than the cache servers of the Fox Sports Go live streaming. Such an increase in the 90th percentile latency due to the Super Bowl was only around 24.9\%. As both 90th percentile latencies were under 100 ms and the latency increase due to Super Bowl was less than 20 ms, we believe that all CDNs had scaled resource provisioning well enough resource to guarantee a quality live streaming for Super Bowl.

%% file: webservice-impact.tex
\section{Super Bowl impact on popular services}
Different from old days when the families got together watching Super Bowl on cable TVs, nowadays users interact more actively online. They might watch the game via live streaming service, browse their social networks when distracted, and even order food online during the game. These services are all provisioned either from their private Cloud or hybrid Cloud.  We want to understand whether the change of user behaviors during Super Bowl are reflected in the performance of these services during Super Bowl. We studied two types of popular services, on-demand video streaming service and social network services.
\subsection{On-demand video streaming service}
\begin{figure}[!htb]
   \includegraphics[width=0.5\textwidth]{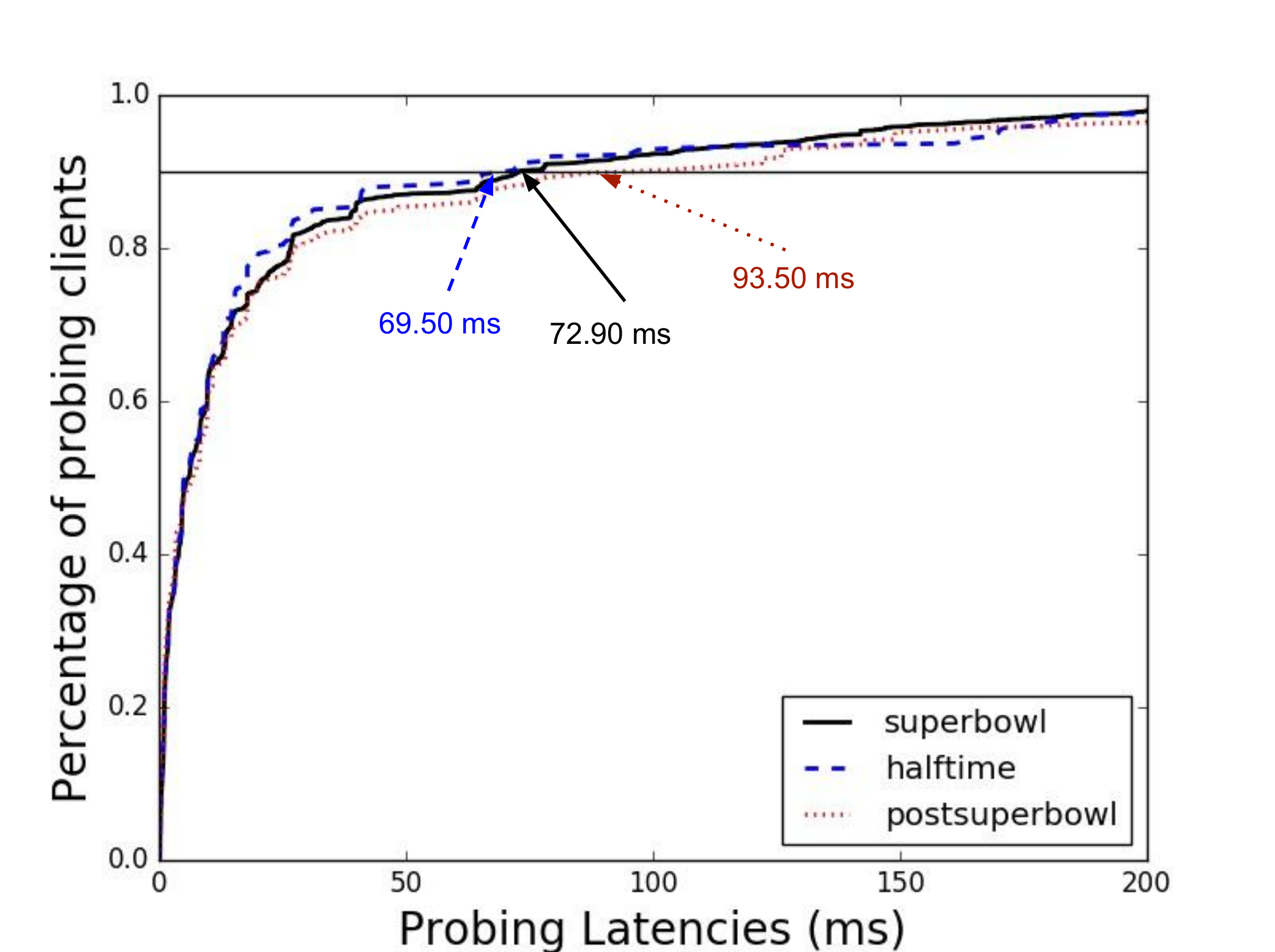}
   \caption{The CDF of probing latencies from PlanetLab agents to Netflix cache server in 2017}
   \label{fig:netflix_CDN_lat_cdf_2017}
\end{figure}

\begin{figure}[!htb]
   \includegraphics[width=0.5\textwidth]{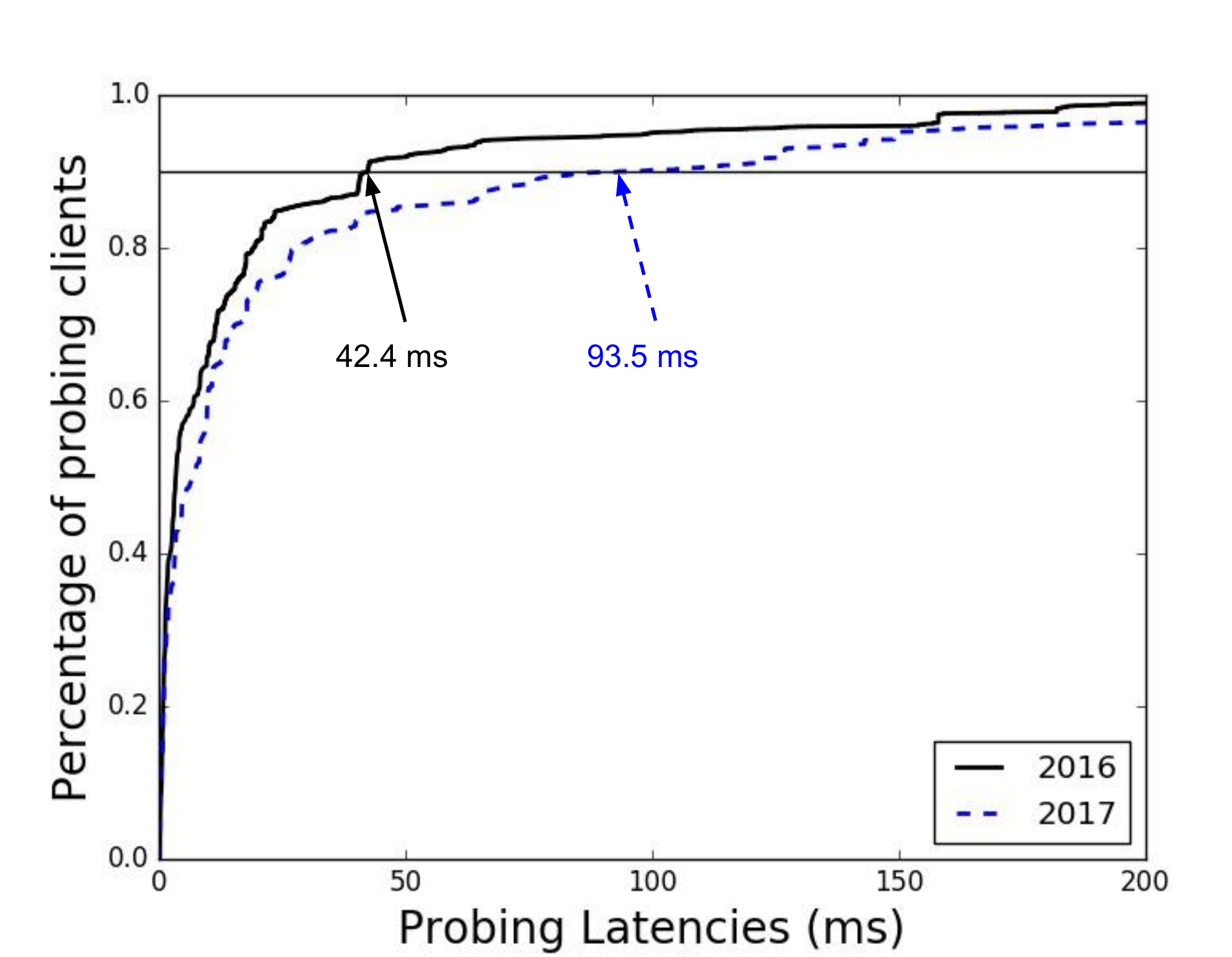}
   \caption{The CDF comparison of probing latencies from PlanetLab agents to Netflix cache server at Sunday nights on 2016 and 2017}
   \label{fig:netflix_CDN_lat_cdf_over_2yrs}
\end{figure}

We probed one of the CDN host (\url{netflix753.as.nflximg.com.edgesuite.net}) used for Netflix video streaming to study the on-demand video streaming service. We let 100 PlanetLab servers probe one Netflix streaming host every 1 minute. We first compared the network latencies from PlanetLab agents to the Netflix host during three periods, the \textit{superbowl}, the \textit{halftime} and the \textit{postsuperbowl} periods in Figure \ref{fig:netflix_CDN_lat_cdf_2017}. We observe that the overall latencies during the \textit{superbowl} and the \textit{halftime} periods were lower than the \textit{postsuperbowl} period, which was at the  Sunday night after Super Bowl LI. However, the difference in latencies were not significant. The ninetieth percentile latencies to the Netflix host are 69.5 ms in the \textit{halftime} period, 72.90 ms in the \textit{superbowl} period, and 93.50 ms in the \textit{postsuperbowl} period. The decrease of 90th percentile latency due to the Super Bowl was only 25.67\%. The Super Bowl can be considered as a relatively quiet period for Netflix. The low overall latencies and the small variations in latencies verified that Netflix had light workload during the game.
From probing the Netflix cache server in the past 2 years (2016 and 2017) at the Sunday night after Super Bowl, as shown in Figure \ref{fig:netflix_CDN_lat_cdf_over_2yrs}, we also find that the latencies to the Netflix cache server increased over the past 2 years. If we compare the 90th percentile latency from PlanetLab agents in US to the same Netflix cache server over the past 2 years, the latency increased from 42.4ms to 93.5 ms. This was a more than 2x increase per year in the 90th percentile latency. It shows that the Netflix probably got a huge increase in traffic at Sunday nights and the latencies were increasing. It also indicates that Netflix did not scale their resource provisioning fast enough to catch their fast increasing user demand. However, regardless of the increasing user demand, Netflix was still able to maintain their 90th percentile latencies below 100 ms.
\subsection{Social networks}
During the Super Bowl game, we had two opposite speculations about users' behaviors. The first one expected users would be fully caught by the game and they would have fewer online activities. The second one expected users would be more active online, discussing about the exciting moments in the game, or even posting their experience online. We used 100 PlanetLab agents in US to probe the top two popular social networks, Twitter and Facebook, to study which speculation was true. 

Our network measurements to Facebook cache server supported the first speculation as shown in Figure \ref{fig:fb_CDN_lat_cdf_over_sbperiods}. The latencies in the \textit{postsuperbowl} period were overall longer than the latencies in the \textit{superbowl} period. We observe at the 70th percentile latency, where there was a big gap among curves of different periods. It shows that the 70th percentile latency in the \textit{postsuperbowl} period was more than 10 ms longer than the 70th percentile latencies in the \textit{superbowl} and the \textit{halftime} periods. It is worth noting that the \textit{halftime} period had higher 85th percentile latency than the \textit{postsuperbowl} period. The high 85th percentile latency in the \textit{halftime} period can be caused by some load spikes in the Facebook cache servers during the halftime show. It might be related to more active social network activities during the halftime show, however we did not have the volume of Facebook posts to verify that.

Our network measurements to Twitter cache server supported the second speculation as shown in Figure \ref{fig:fb_CDN_lat_cdf_over_sbperiods}. We see that the overall latencies to Twitter cache server did not vary a lot among different periods. However, if we observe the 90th percentile latency, we can see that the \textit{superbowl} and the \textit{halftime} curves are on the right of the \textit{postsuperbowl} curve. The Super Bowl game increased the 90th percentile latency from 137.00ms to 161.00 ms. The half time show increased it further to 163.00 ms. The increase can be caused by the load spike on Twitters, which might be related to more active usage of Twitter during the Super Bowl game. We can see that at the 80th percentile, the latency was around 50 ms and there were only slight differences among three periods. Namely, the majority of users (80\% of users) would not be able to notice the latency difference due to the Super Bowl game.

\begin{figure}[h]
   \includegraphics[width=0.5\textwidth]{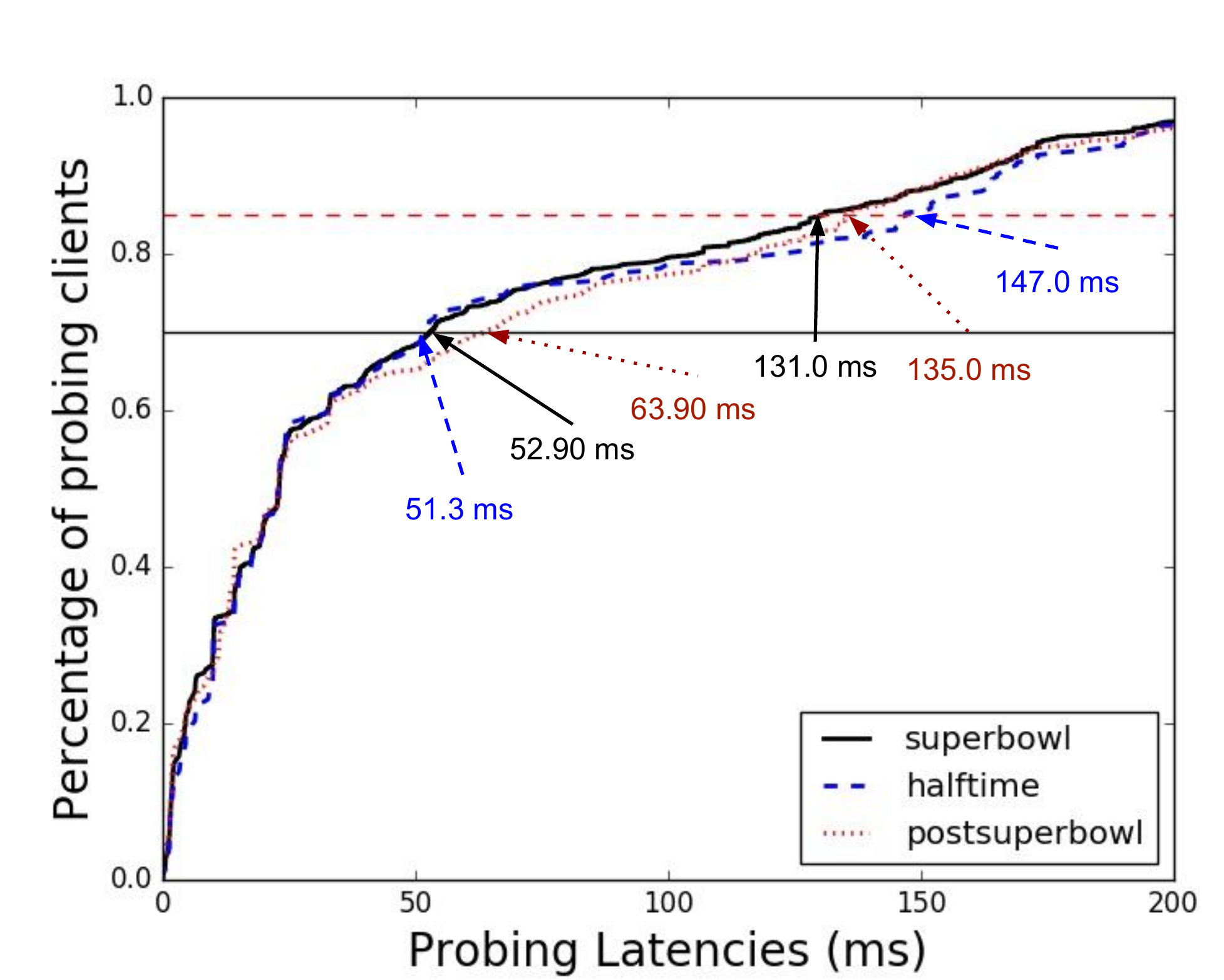}
   \caption{The CDF of probing latencies from PlanetLab agents to Facebook cache server in 2017}
   \label{fig:fb_CDN_lat_cdf_over_sbperiods}
\end{figure}
\begin{figure}[!htb]
   \includegraphics[width=0.5\textwidth]{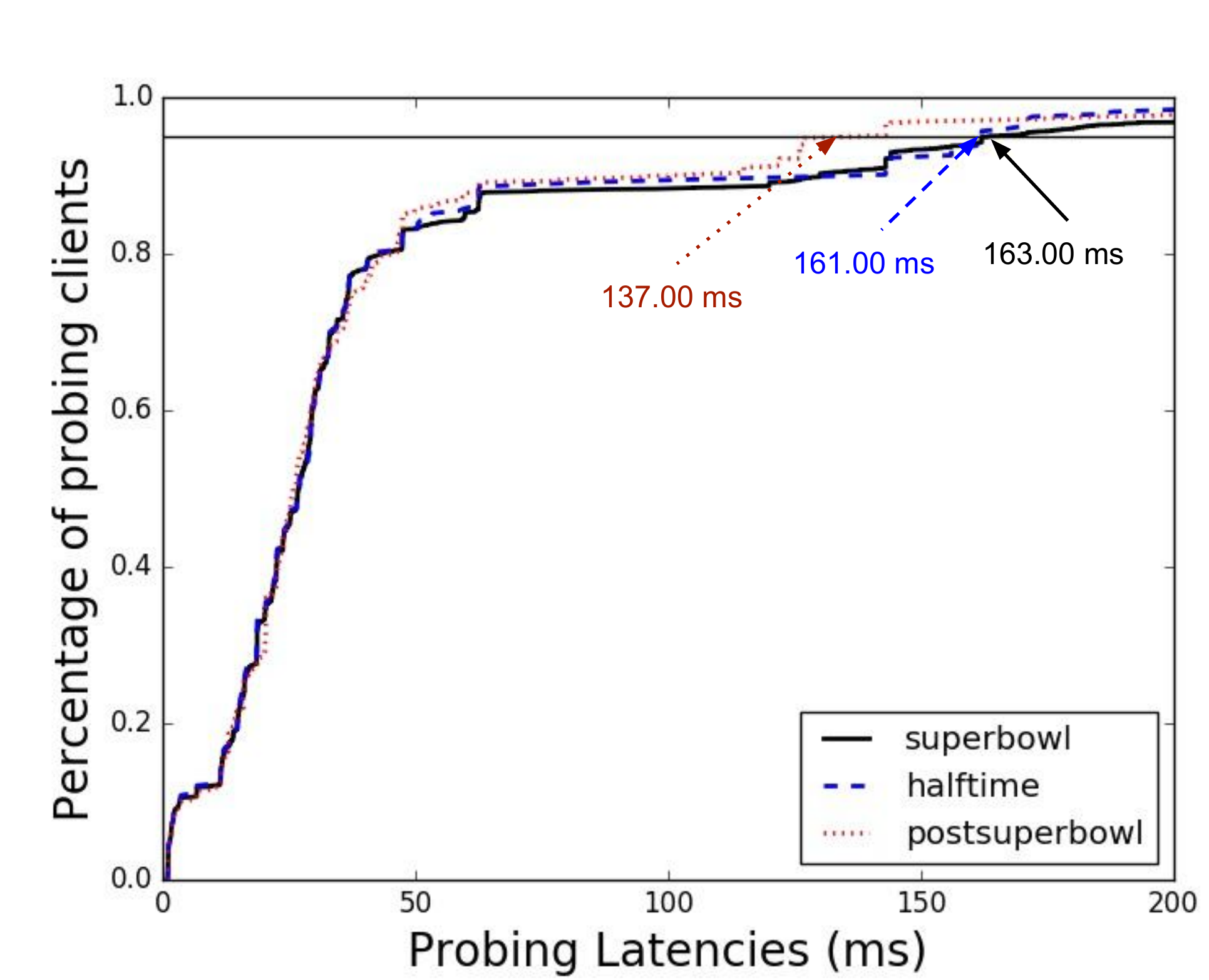}
   \caption{The CDF of probing latencies from PlanetLab agents to Twitter cache server in 2017}
   \label{fig:fb_CDN_lat_cdf_over_sbperiods}
\end{figure}

%% file: conclusion.tex
\section{Conclusions}
In this paper, we probed various types of Cloud systems to study the impact of the Super Bowl games from 2015 to 2017. More specifically, we probed the Cloud infrastructures, the CDNs, and popular web applications to study how the Super Bowl games and the half-time shows impact these systems. We obtained the following insights from the study: 1) we observed an increase in latencies to access to Cloud VMs during the Super Bowl games in recent years, which is very likely related to the increasing Internet traffic due to the increasing online activities; 2) Both the intra-data center and inter-data center network latencies increase in the Cloud during the Super Bowl. However, such impact is insignificant to the performance of the Cloud Infrastructure as a Service (IaaS); 3) when Fox Sports Go service set up multiple CDNs for Super Bowl live streaming, including Akamai, Level 3 and Limelight in 2017, latencies to these CDN cache servers varied across providers. However, CDN providers were able to allocate resources appropriately, as the increase of 90th percentile latency due to the Super Bowl game was less than 20 ms; 4) the Super Bowl game also affected popular web services differently. More specifically, it reduced the latencies to cache servers of Netflix  and Facebook but increased the latencies to cache servers of Twitter. Our results show that although the Super Bowl affected the latencies to these Cloud systems hosing different web services, the  current  Cloud  system  is  able  to  provision elastic capacity to minimize such impact caused by even  a massive event. 